\documentclass[twocolumn]{jpsj2}

\def \virg{\;\;,}
\def \point{\;\,.}

\usepackage{amsmath}

\title{%
Pressure-Induced Zero-Gap Semiconducting State in 
 \\  Organic  Conductor  $\alpha$-(BEDT-TTF)$_2$I$_3$ Salt
}

\author{%
Shinya Katayama, Akito Kobayashi and   Yoshikazu Suzumura 
}

\inst{%
Department of Physics, Nagoya University, Nagoya 464-8602
}

\recdate{December 5, 2005}

\abst{%
We show a zero-gap semiconducting (ZGS) state 
 in the quasi-two-dimensional organic conductor $\alpha$-(BEDT-TTF)$_2$I$_3$ 
  salt, 
  which emerges  under uniaxial pressure 
   along the $a$-axis (the stacking axis of the  BEDT-TTF molecule).
       The ZGS state is the state in which 
     a Dirac cone with the band spectrum of a linear dispersion 
        exists around the Fermi point  
         connecting an unoccupied (electron) band  
          with an  occupied (hole) band. 
The spectrum exhibits a large anisotropy in velocity, 
  which depends on the direction from the Fermi point.  
     By varying   the magnitude of several  transfer energies 
   of a tight-binding model with four sites  per unit cell,
          it is shown that the ZGS state 
     exists   in a wide pressure range,  and 
     is attributable to the large anisotropy of the transfer energies 
       along the stacking axis. 
}

\kword{organic conductor, BEDT-TTF, quarter-filling, 
tight-binding model, zero-gap semiconductor, Dirac cone, 
contact point, $\alpha$-(BEDT-TTF)$_2$I$_3$, 
}

\newcommand{\vep}{\varepsilon}

\newcommand{\jpsj}[3]{: J. Phys. Soc. Jpn. {\textbf{#1}} ({#2}) {#3}}
\newcommand{\prb}[3]{: Phys. Rev. B \textbf{#1} ({#2}) {#3}}

\begin{document}
\maketitle

\section{Introduction}\label{intro}
Quasi-two-dimensional organic conductors, BEDT-TTF 
(bis(ethylene-dithio)tetrathiafulvalene) salts, exhibit several 
electronic states, such as 
       Mott insulator, charge ordering and superconductivity, 
\cite{Ishiguro_1998,Hseo1}
       under the variation of temperature, 
         hydrostatic pressure or uniaxial strain. 
The electronic states of these BEDT-TTF (ET) salts originate from  
  $\pi$-electrons with the several transfer energies and 
  the strong Coulomb interaction acting on-site and between nearest-neighbor sites.
Among these  salts, unusual phenomena have been observed in 
  the $\alpha$-(ET)$_2$I$_3$ ($\alpha$ phase of BEDT-TTF tri-iodide) salt, 
\cite{Bender} 
 which consists of four molecules per unit cell  
   with a 3/4-filled  band (i.e., quarter-filled band of the hole). 
Under ambient pressure (A.P.),  
 the $\alpha$-(ET)$_2$I$_3$ salt exhibits the metal-insulator transition 
    at $T_{MI}=135$ K, which is followed by charge ordering 
\cite{Rothaemel,Kajita,Tajima_2000,Takano,Wojciechowskii,Takahashi_SM_2003} 
  along the $b$-direction perpendicular to 
   the stacking axis ($a$-axis). 
When a uniaxial strain at 2 kbar is applied  along the $a$-direction, 
  the salt  undergoes the superconducting transition at 7 K.
  \cite{Maesato,Tajima_2002}
Under high  pressures, the $\alpha$-(ET)$_2$I$_3$ salt shows exotic properties, that 
 are consistent with the narrow gap semiconducting (NGS) state.
\cite{Kajita,Tajima_2000} 
At 20 kbar of the  hydrostatic pressure, 
 resistivity becomes metallic and is almost independent of   
   temperature between 300 K and 1.5 K, 
 although carrier density, $n$, decreases 
 from $10^{21}$ cm$^{-3}$ (300 K) to $10^{15}$ cm$^{-3}$ (1 K). 
It has been claimed that 
  such a temperature dependence may originate from  
     the NGS state with a gap $E_g\sim 1$ meV. 
 \cite{Tajima_2000}
 There exists a difference in the effect of uniaxial pressure   
 between the $a$-direction ($P_a$) and  the  $b$-direction ($P_b$).
\cite{Tajima_2002}
 For a compression along the $a$-direction, 
        the pressure dependence of $T_{MI}$ is small
         and the NGS state is likely to be realized for $P_a > 5$ kbar. 
For the uniaxial strain along the $b$-direction, 
  the metal-insulator transition is suppressed noticeably, 
   whereas neither superconducting  transition nor 
      NGS state occurs at low temperatures.

The ordered state in $\alpha$-(ET)$_2$I$_3$ has been investigated theoretically; 
 a charge ordering state with the horizontal pattern was 
     obtained using the Hartree-Fock approximation 
\cite{Kino_JPSJ_ET,Kino_JPSJ_ET2,Seo_JPSJ,hotta} 
  and the  superconducting state 
    in the presence of the charge ordering 
   was clarified in terms of the random phase approximation
\cite{Kobayashi_JPSJ,Kobayashi_JPSJ2}.   
It is noted that such a  superconducting state occurs  
 just before the onset of the  NGS state under the $a$-axis pressure. 
   However, the origin of the NGS state  is  not yet clear,  
  although only the uniaxial pressure along the 
  $a$-axis leads to the NGS state in the $\alpha$-(ET)$_2$I$_3$ salt.

In the present paper, we examine the pressure dependence of 
 the band structure 
  of the $\alpha$-(ET)$_2$I$_3$ salt within the tight-binding model 
  in order to understand the NGS state, which emerges and remains 
     in a wide range of  the uniaxial pressure along the $a$-axis. 
In \S 2, we study the pressure dependence by assuming an 
   extrapolation formula for the pressure dependence of  
   transfer energies of the $\alpha$-(ET)$_2$I$_3$ salt. 
We employ  the  data 
     obtained from the  X-ray diffraction experiment
      under pressures along   both the $a$-axis and the $b$-axis.  
\cite{Mori,Kondo,Kondo_2} 
In \S 3, it is shown that 
 the present organic conductor  under high pressures 
   exhibits  a zero-gap semiconducting (ZGS) state with a Dirac-cone-like 
     linear dispersion  instead of  the NGS state. 
The origin of the ZGS state under pressure is analyzed by varying parameters of 
transfer energy in \S4.   
We also investigate the ZGS state in \S5 by using a reduced model  
   with   two sites in the  unit cell. 
Section 6 is devoted to the summary and discussion. 

\section{Formulation}

\begin{figure}[htbp]
\begin{center}
\includegraphics[width=6.0cm]{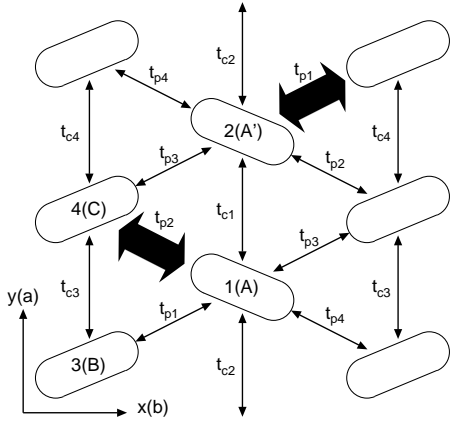}
\caption{
Conducting plane of $\alpha$-(ET)$_2$I$_3$, where there are  four ET molecules in the unit cell 
and the bond represents eight transfer energies $(t_{p1},\cdots,t_{p4},t_{c1},\cdots ,t_{c4})$. 
The $x(y)$-axis corresponds to the $b(a)$-axis in ref.~\citen{Mori}.
Thick arrows denote the dimerization in a model, which is considered in \S\ref{site2}.
}
\label{unitcell}
\end{center}
\end{figure}

In order to describe conduction electrons for the $\alpha$-(ET)$_2$I$_3$ salt, 
we start with a two dimensional lattice model, which 
 consists of four ET molecules in the unit cell 
  as shown in Fig.~\ref{unitcell}.
The Hamiltonian is given by  
\begin{align}
H&=\sum_{
\mathrm{n.n.}
}t_{i\alpha :j\beta}
c^{\dag}_{i\alpha}c_{j\beta} \virg 
\label{h4}
\end{align}
where $t_{i\alpha :j\beta}$ is the transfer energy from 
the $(j,\beta)$ site to the nearest-neighbor 
(n.n.) $(i,\alpha)$ site. 
In eq.~(\ref{h4}),  $i$ and $j$ $(=1,\cdots,N_L)$ denote 
  the unit cells of the square lattice. 
The indices $\alpha$ and $\beta$ denote 
 the four molecules in the unit cell, 
 where 1, 2, 3 and 4 
 correspond  to A, A', B and C in ref.~\citen{Mori}, respectively. 
$c_{i\alpha}$ is the annihilation operator of the conduction electron 
 at the $(i,\alpha)$ site. 
Using the  Fourier transformation, 
$c_{i\alpha}={N_L}^{-1/2}
\sum_{\mib{k}}e^{i\mib{k}\cdot\mib{r}_i}c_{\mib{k}\alpha}$, 
eq.~(\ref{unitcell}) is rewritten as 
\begin{align}
H&=\sum_{\mib{k},\alpha,\beta}\vep_{\alpha\beta}(\mib{k})
c^{\dag}_{\mib{k}\alpha}c_{\mib{k}\beta} \virg 
\label{h4f}
\end{align}
where
\begin{align}
\vep_{\alpha\beta}(\mib{k})&=
\dfrac{1}{N_L}
\sum_{i,j}t_{i\alpha :j\beta}e^{-i\mib{k}\cdot(\mib{r}_i-\mib{r}_j)} \virg  
\label{ep}
\end{align}
and $t_{i\alpha :j\beta}$ represents eight transfer energies 
  given by 
 $t_{p1},\cdots,t_{p4},t_{c1},$ $\cdots,t_{c4}$  as shown in  Fig.~\ref{unitcell}.  
 In eq.~(\ref{h4f}),
   $\vep_{\alpha\beta}(\mib{k})$ is expressed as 
\begin{align}
\vep_{\alpha\alpha}(\mib{k})&=0\quad(\alpha=1,\cdots,4) \virg
 \nonumber \\
\vep_{\alpha\beta}(\mib{k})&=\vep_{\beta\alpha}^{*}(\mib{k}) \virg 
\nonumber\\
\vep_{12}(\mib{k})&=t_{c1}+t_{c2}e^{-ik_y} \virg 
\nonumber\\
\vep_{13}(\mib{k})&=t_{p1}-t_{p4}e^{ik_x} \virg 
\nonumber\\
\vep_{14}(\mib{k})&=t_{p2}-t_{p3}e^{ik_x} \virg
\nonumber\\
\vep_{23}(\mib{k})&=t_{p4}e^{ik_y}-t_{p1}e^{i(k_x+k_y)} \virg 
\nonumber\\
\vep_{24}(\mib{k})&=t_{p3}-t_{p2}e^{ik_x} \virg 
\nonumber\\
\vep_{34}(\mib{k})&=t_{c3}+t_{c4}e^{-ik_y} \virg 
\label{h42}
\end{align}
where $k_x$ is replaced by $k_x + \pi$, and 
the suffix  $x(y)$ corresponds to the $b$($a$)-axis.\cite{Mori}
The chemical potential 
determined by the 3/4 filling is set to zero 
 in order to take  the Fermi energy  as the origin of the band spectrum.

For  the pressure dependence of 
  transfer energies   $t_{i\alpha:j\beta} = t_A, t_A' $ 
($A =c1,\cdots,c4,$ $p1, \cdots, p4$)
   of $\alpha$-(ET)$_2$I$_3$,   
      we employ an extrapolation formula given by
    \cite{Kobayashi_JPSJ}   
\begin{align}
 \label{transfer_P}
 t_A(P_a)= t_A ( 1 + K_A P_a)\virg \\
  t_A'(P_b)= t_A ( 1 + K_A' P_b) \virg
 \label{transfer_PB}
\end{align}
  where 
  $t_A(P_a)$ ($t_A'(P_b)$) represents 
  the  transfer  energy under the pressure $P_a$ ($P_b$) along the $a$-axis 
   ($b$-axis).  
  Hereafter, the units of pressure and transfer energy are taken as 
   kbar and eV, respectively.   
 By using  the data of  $\alpha$-$\mathrm{(ET)_2 I_3}$ salt 
  for ambient pressure ($P_a=P_b=0$)\cite{Mori} 
   and those  for   $P_a = 2$ and  $P_b = 3$
   at room temperature, 
 \cite{Kondo,Kondo_2}
the above coefficients  are estimated as 
$t_{p1}=0.140$, 
$t_{p2}=0.123$, 
$t_{p3}=-0.025$, 
$t_{p4}=-0.062$, 
$t_{c1}=0.048$, 
$t_{c2}=-0.020$, 
$t_{c3}=-0.028$, 
$t_{c4}=-0.028$ 
at ambient pressure and 
$K_{p1}=0.011$,
$K_{p2}=0.0$,
$K_{p3}=0.0$,
$K_{p4}=0.032$,
$K_{c1}=0.167$,
$K_{c2}=-0.025$, 
$K_{c3}=0.089$ and 
$K_{c4}=0.089$ 
($K'_{p1}=0.024$,
$K'_{p2}=0.031$,
$K'_{p3}=0.053$,
$K'_{p4}=0.022$,
$K'_{c1}=0.042$,
$K'_{c2}=0.133$, 
$K'_{c3}=0.167$ and 
$K'_{c4}=0.167$) 
for pressures along the $a$-axis ($b$-axis). 
As seen from $K_A$,
 the large $K_{c1}$ suggests  that the  pressure dependence 
 along the $a$-axis  is mainly determined by  $t_{c1}$.

\section{ZGS State}\label{chapngs}
In this section, we focus on the electronic states under 
the pressure along the $a$-axis, i.e., $P_a$. 
Since there are four sites in the unit cell, 
 there are four energy bands which are obtained by diagonalizing eq. (\ref{h4f}). 
The first band with the highest energy 
  is almost empty and the second one is almost filled
   due to 3/4-filling.
Whereas 
  there is a small electron pocket 
  close to the X$(\pm \pi,0)$ point 
    and a hole pocket  close to the Y$(0,\pm \pi)$ point at ambient pressure,
    \cite{Tajima_2000}
     these pockets are reduced at $P_a$ = 2 
     \cite{Kondo,Kondo_2}
      and they are further reduced to a point  
        in a wide range of higher pressures.
        \cite{Kobayashi_JPSJ,Kobayashi_JPSJ2} 
  Actually from eq.~(\ref{transfer_P}), 
  it is found that  such an unusual state is stable 
  for $P_a > 3$.

\begin{figure}[htbp]
\begin{center}
\includegraphics[width=8.0cm]{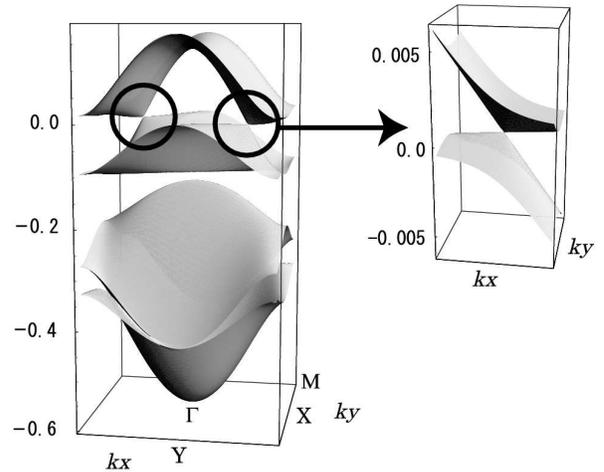}
\caption{Band dispersion of $\alpha$-(ET)$_2$I$_3$ at $P_a=4$ kbar 
in first Brillouin zone (left panel) and enlarged one 
around contact point at $\mib{k}^0=(0.602\pi,-0.353\pi)$ (right panel), 
where Fermi energy is taken as origin.
}
\label{band1}
\end{center}
\end{figure} 
Figure~\ref{band1} shows the energy  band spectrum  
  of $\alpha$-(ET)$_2$I$_3$ at $P_a$ = 4  (left panel) in the plane of $k_x$ and $k_y$, 
   which  is scaled by the inverse of the lattice constant. 
 The first band is unoccupied, 
  whereas the second band is occupied. 
  It should be noted that 
  the first and second bands touch 
   at the points $ \pm \mib{k}^0$ as shown with the circle. 
  This  contact point becomes   the Fermi point 
 due to 3/4-filling. 
Actually,  
the matrix elements of eq.~(\ref{h42}),
corresponding to 
 $\mib{k}^0 = (0.602\pi,-0.353\pi)$,
  are   
 $
   \vep_{12} = 0.072 -  0.016i, ~~ 
   \vep_{13} = 0.124 +  0.066i, ~~ 
   \vep_{14} = 0.115 +  0.024i, ~~ 
   \vep_{23} = -0.135 -  0.040i, ~~ 
   \vep_{24} = 0.014 -  0.117i$ and 
$\vep_{34} = -0.055 -  0.034i$. 
The four  band energies are given as 
    0, 0, -0.210,  -0.477,
  in which the Fermi energy is given 
   by $\vep_F=0.172$.
  The existence of such a point 
   implies that 
     the hole and electron bands are degenerate at the point. 
Since the point has no relation to the symmetry,
  such a degeneracy occurs accidentally 
   as first pointed out by Herring. 
 \cite{Herring}

\begin{figure}[htbp]
\begin{center}
\includegraphics[width=8.0cm]{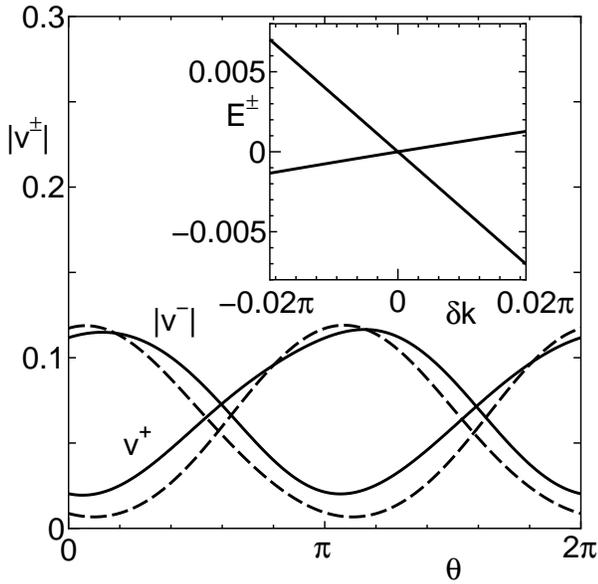}
\caption{
Angular dependence of velocity 
around contact point, i.e., Dirac cone 
 at $P_a$ =10  (solid line)  and 5  (dashed line), 
 where 
    $\theta$ (horizontal axis) denotes angle between 
      $\delta \mib{k} (=\mib{k}-\mib{k}^0 )$
     and the $k_x$-axis.
 The inset shows  $ E^{\pm}(\delta \mib{k})$ at   $P_a=10$, 
 where $\delta k >0$ and $<0$ correspond to $\theta$  = 0 and 
 $\pi$, respectively. 
}
\label{velocity}
\end{center}
\end{figure}

It should be noted in 
  the  enlarged figure  (right panel of Fig.~\ref{band1}) that 
   a linear dispersion is present around 
      the contact point   $\mib{k}^0$.  
 Such a state  always 
  exists for $P_a > 3$. 
    The dispersion  close to the point can be expressed as  
\begin{align}
 E^{\pm}(\delta \mib{k}) =  v^{\pm}(\delta \mib{k})  | \mib{k} - \mib{k}^0 |
 \virg      
\label{contact_spect}
\end{align}
 where  $ \delta \mib{k} = \mib{k} - \mib{k}^0$, 
 and  $v^+(\delta \mib{k})$ ($v^-(\delta \mib{k})$) is 
 the  velocity for the electron (hole).
  We call such a state the Dirac cone, 
  according to the literature.
  \cite{Wolff,Kohno,Abrikosov,Lukyancuk}
 In Fig.~\ref{velocity}, 
 the velocity  $v^{\pm}(\delta \mib{k})$   is shown 
  as a function of  
    $\theta$, which is the angle between 
    the vector  $ \delta \mib{k}$
     and the $k_x$-axis. 
 A noticeable angular variation in velocity  is seen 
  with the relation: 
\begin{align}
\label{cone_velocity1}
   v^+(\delta \mib{k}) & \not=-  v^-(\delta \mib{k}) \virg  \\
 v^+(\delta \mib{k})  & =-  v^-( - \delta \mib{k}) \virg 
\label{cone_velocity2}
\end{align}
  within the numerical accuracy of the present calculation. 

\begin{figure}[htbp]
\begin{center}
\includegraphics[width=8.0cm]{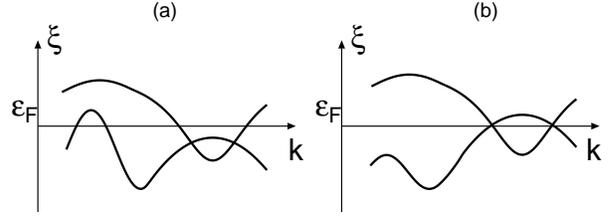}
\caption{
Schematic band dispersion $\xi$ of metallic state at low pressure (a) 
 and that of ZGS state at  high pressure (b), 
 where wave number  $\mib{k}$ (horizontal axis) 
  is chosen so as to  include contact point. 
}
\label{band2}
\end{center}
\end{figure} 
  For both low and high pressures,  
  the schematic band dispersion
     is shown  in   Fig.~\ref{band2}, 
      where the crossing in the band spectrum    
     denotes  the contact point which is located  at the bottom of the electron band. 
At low pressures (Fig.~\ref{band2}(a))  with  $P_a \lesssim 3$, 
 one finds a metallic state where two contact points at 
   $\mib{k}^0$ and $-\mib{k}^0$ exist below the Fermi energy 
    $\vep_{F}$. 
  However, at high pressures (Fig.~\ref{band2}(b)) with $P_a \gtrsim 3$, 
  the contact point is located  just at the Fermi energy, i.e., 
   the realization of the ZGS, 
  which is retained 
     due to the 3/4-filling for a wide pressure range. 
 The ZGS state shows up in  the 
    density of states $D(\omega)$ close to  the Fermi energy as 
 \cite{Kobayashi_JPSJ}  
\begin{align}
D(\omega)\propto |\omega| \virg  
\label{linear}
\end{align}
where $\omega = 0$ corresponds to the Fermi energy.
Since $D(\omega)$ vanishes at  $\omega=0$, 
  we call the state with eq.~(\ref{linear}) the ZGS state 
    instead of the NGS state.
  \cite{Tajima_2000,Kobayashi_JPSJ}
  Although   the contact point  still remains 
   in the metallic state  at low pressure,  
  the corresponding cusp in the density of states 
      is reduced and becomes invisible for $P_a \lesssim 2.5$.

\begin{figure}[htbp]
\begin{center}
\includegraphics[width=7.0cm]{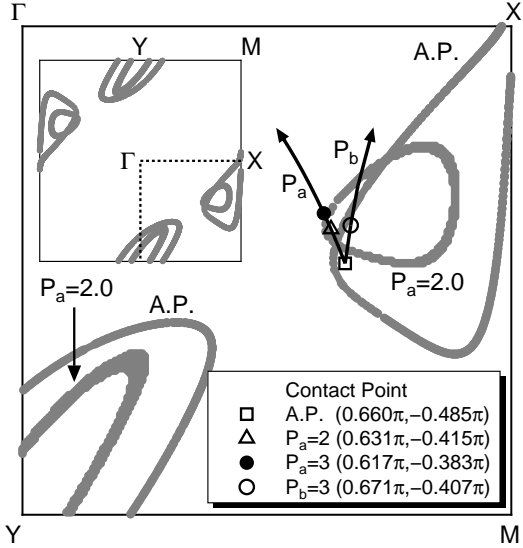}
\caption{
Fermi surface (FS) of $\alpha$-(ET)$_2$I$_3$ at ambient pressure and $P_a=2$, 
 where main (inset) figure corresponds to quarter (full) part of 
 Brillouin zone. 
 The arrow $P_a$ denotes the pressure dependence  of the contact point
   $\mib{k}^0$ for $ 0 < P_a < 10$,
  whereas the arrow $P_b$ corresponds to  $\mib{k}^0$   for $ 0 < P_b < 10$. 
  The square, triangle, and filled circle represent 
   $\mib{k}^0$  for  $ P_a =$ 0, 2, and 3 respectively, 
   while the open circle  denotes that for  $ P_b =$ 3. 
}
\label{etfscp}
\end{center}
\end{figure}

In Fig.~\ref{etfscp}, 
 the pressure dependence of the point contact, $\mib{k}^0$, 
  is examined, where  
  the Fermi surface is shown for 
   ambient pressure and  $P_a =2$.  
With increasing pressure, the area of the Fermi surface is reduced  
 as seen from  the electron (hole) pocket  around the X (Y) point.
 \cite{Kondo}
 The  symbols on the arrow $P_a$ represent   the location  of  contact  points 
 at $P_a =$ 0, 2 and 3, respectively. 
  The arrow  $P_a $, which  denotes the trajectory of  $\mib{k}^0$
    from $P_a  =$ 0  to    $P_a =$ 10,
   extends  to the direction of the $\Gamma$ point 
    with increasing $P_a$. 
 The contact point is located   below the Fermi surface  for $P_a < 3$ and 
   just on the Fermi surface   for $3 <  P_a < 10$. 
 The arrow $P_b$, which  corresponds to that of $\mib{k}^0$ 
  for $ 0 < P_b < 10$, 
   is discussed  in \S 6. 
    
\section{Stability  of Contact Point}\label{parac}
In order to examine the stability  of the  contact point
against the change in parameters, 
 we  calculate the energy band spectra  
 by varying  magnitudes  of the transfer energies
  from those of $\alpha$-(ET)$_2$I$_3$. 
 Since the difference between $t_{c1}$ and $t_{c2}$ is large  for $\alpha$-(ET)$_2$I$_3$, 
   we employ a model, in which $t_{c1}\neq t_{c2}$ and 
    the remaining parameters  are chosen as  
\begin{align}
t_{p1}=t_{p2}=1 \virg  \;\; t_{p3}=t_{p4} \point 
\end{align} 
Thus, we examine a simplified model with independent parameters, 
   $t_{c1}$, $t_{c2}$, $t_{p3}$ and $t_{c3}$. 
\begin{figure}[htbp]
\begin{center}
\includegraphics[width=7.0cm]{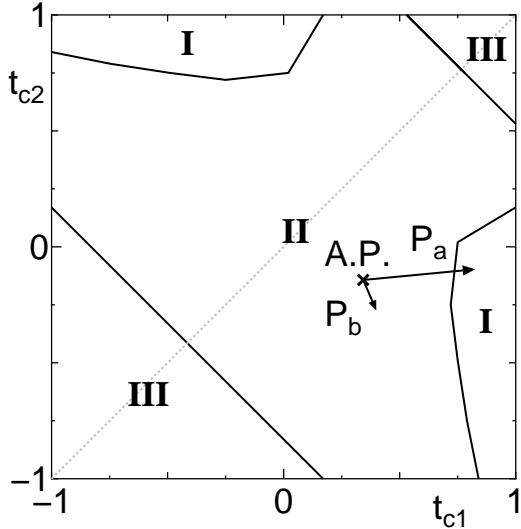}
\caption{
Phase diagram on plane of $t_{c1}$ and $t_{c2}$
  for $t_{p3}=t_{p4}=t_{c3}=t_{c4} = -0.25$,  
 where transfer energies $t_{c1}$ and $t_{c2}$ are 
   also  normalized by $t_{p1} (=t_{p2})$. 
The arrow denotes the  the variation in 
 $(t_{c1},t_{c2})$ 
   estimated from 
     eq.~(\ref{transfer_P}) 
 for $0< P_a < 10$ and $0 < P_b <5$, respectively, 
  where  the cross corresponds to  ambient pressure. 
}
\label{c1c21}
\end{center}
\end{figure} 

Figure~\ref{c1c21} shows phase diagram in the plane of 
$t_{c1}$ and $t_{c2}$ for fixed $t_{p3}$ $(=t_{p4}=t_{c3}=t_{c4})=-0.25$. 
 The  contact points exist 
      in regions  I and II, which correspond  to 
        Figs.~\ref{band2}(b) and 4(a), respectively.
  Region I exhibits the ZGS state, 
   where  the  point becomes equal to the Fermi point. 
Region II corresponds to the metallic state 
    with  the point located below the Fermi energy.  
In Region III,  the contact point disappears 
  and the conventional metallic state is found. 
 The boundary between regions II and III is given by 
\begin{align}
t_{c1}+t_{c2}=r_1, ~~~~~t_{c1}+t_{c2}=r_2 \virg 
\label{cont}
\end{align}
where $r_1=1.53$ and $r_2=-0.83$. 
Such a straight line of the boundary is due to 
  the choice of the parameter 
     $t_{p3}=t_{p4}=t_{c3}=t_{c4}$. 
 When parameters $t_{c1}$ and $t_{c2}$ take the value on the boundary of 
 $r_1$ ($r_2$), 
  the contact point is given by the $\Gamma$ (X) point 
    on the $k_x$-$k_y$ plane in Fig.~\ref{etfscp}. 
Region I corresponds to a large $t_{c1}$ and a negative $t_{c2}$  
 (or a large $t_{c2}$ and a negative $t_{c1}$). 
Note that the case of $t_{c1}=t_{c2}$ (dotted line in Fig. \ref{c1c21}) gives 
 the metallic state with a special situation 
    where the first band has the same energy as the second band 
      on the line $k_y=\pm\pi$ in the  Brillouin zone. 
The contact point disappears on this line. 
  The van Hove singularity in the density of states exists 
   near the Fermi energy 
    when the parameters in region I move close to region II. 
When $|t_{p3}|(=|t_{p4}|)$ and $|t_{c3}|(=|t_{c4}|)$ are decreased, 
   the area of region II is reduced. 

 Here, we comment on  the pressure dependence of the contact point
 in Fig.~\ref{c1c21}. Here the arrow denotes  
   the variation of  $t_{c1}$ and $t_{c2}$
    under the uniaxial pressures, $P_a$ and $P_b$. 
  From the comparison of these two arrows, it is found that 
       the increase in $P_a$  leads to 
       the ZGS state but that in $P_b$ is away from the ZGS state
       since the increase in $t_{c1}$ is large enough for $P_a$ 
         but is small for $P_b$.

\section{Reduced Model for ZGS State}\label{site2}
The condition for the ZGS state with contact points depends on 
 the transfer energies, although some  of them are redundant.  
   In \S\ref{chapngs}, the ZGS state was 
   obtained for the model 
     shown in Fig.~\ref{unitcell}. 
It consists of four sites in the unit cell. 
However, the model is complicated to examine the ZGS state analytically. 
In order to understand clearly the ZGS state, 
   we reduce 
   further the number of parameters to obtain a model 
      with  two sites per  unit cell. 
  \cite{Sugawara}
There are other reduced models reproducing  the energy band spectrum  
 of   (ET)$_2$X,  
 \cite{Kino_JPSJ_ET2,hotta}
 which will be discussed in the next section.

\begin{figure}[htbp]
\begin{center}
\includegraphics[width=7.0cm]{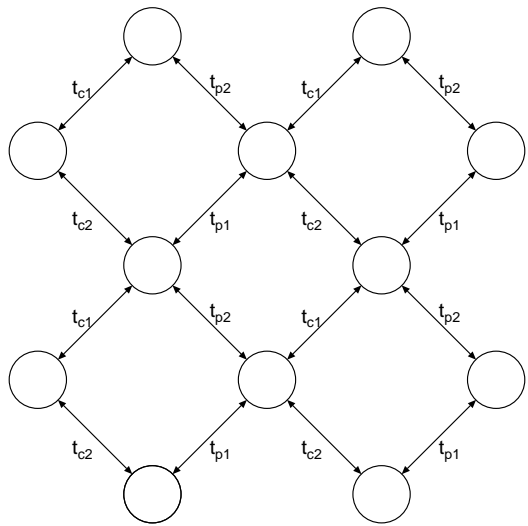}
\caption{
Reduced model with anisotropic square lattice, which is obtained from Fig.~\ref{unitcell}. 
Here two sites are present in unit cell.
}
\label{et2}
\end{center}
\end{figure}

First, we ignore four transfer energies, $t_{p3},t_{p4},t_{c3}$ and $t_{c4}$, 
 in Fig.~\ref{unitcell}, 
 since the ZGS state with  contact points exists  
    even for $t_{p3}=t_{p4}=t_{c3}=t_{c4}=0$.
  However, such a choice is not realistic for $\alpha$-(ET)$_2$I$_3$ at $P_a = 0$  
   due to the absence of  the metallic state as shown later. 
Next, for the simplicity of the calculation, 
 we reduce the number of molecules by 
 considering the dimerization 
 for the bond  $t_{p2}$   between  1 and 4 sites  
  (and also that for  $t_{p1}$   between 2 and 4 sites), 
 which is shown  by the thick arrow 
   in Fig.~\ref{unitcell}. 
With such a procedure, 
 Fig.~\ref{unitcell} can be replaced by   Fig.~\ref{et2}, where we obtain 
 an  anisotropic square lattice consisting  of two sites per unit cell.
  Thus, the reduced model is described by 
  the  Hamiltonian
\begin{align}
H&=\left(
\begin{array}{cc}
0&T
\\
T^{*}&0
\end{array}
\right) \virg \nonumber\\
T&=t_{p1}+t_{p2}e^{ik_y}-t_{c1}e^{i(k_x+k_y)}-t_{c2}e^{ik_x} \virg 
\label{twosites}
\end{align}
with a  half-filled band. 
Hereafter, we set $t_{p1}=t_{p2} (=1)$ for simplicity. 
Two bands are degenerate at $\mib{k}^0$   
 when the off-diagonal component of eq.~(\ref{twosites}) vanishes. 
The wave vector, at which the degeneracy occurs,  can be obtained 
 analytically as
\begin{align}
k_x^0&=\pm 2\tan^{-1}
\left( \sqrt{\dfrac{2-t_{c1}-t_{c2}}{2+t_{c1}+t_{c2}}}\right) \virg 
 \nonumber \\
k_y^0&=\pm 2\tan^{-1}
\left(\dfrac{\sqrt{(2-t_{c1}-t_{c2})(2+t_{c1}+t_{c2})}}{t_{c2}-t_{c1}}\right)
 \point 
\label{k0xk0y}
\end{align}
Equation~(\ref{twosites}) gives 
 a symmetric linear dispersion of the Dirac cone
 with $v^+(\delta \mib{k})  =  -  v^-(\delta \mib{k})$. 
It is slightly different from  eq.~(\ref{contact_spect}).

\begin{figure}[htbp]
\begin{center}
\includegraphics[width=7.0cm]{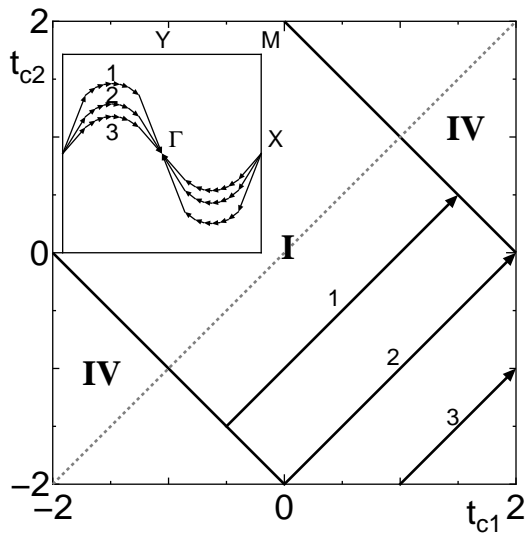}
\caption{
Phase diagram on $t_{c1}$-$t_{c2}$ plane 
for the reduced model given by  Fig.~\ref{et2}. 
Region I is the same as that in 
Fig.~\ref{c1c21}, whereas 
 region IV corresponds to the insulating state.  
The dotted line shows $t_{c1}=t_{c2}$, where 
the upper and lower bands are 
degenerate on the zone boundary. 
The inset shows the trajectory of 
two contact points, when 
 $t_{c1}$ and $t_{c2}$ are varied as the arrow in the main frame. 
}
\label{c1c22site}
\end{center}
\end{figure} 

Figure~\ref{c1c22site} shows  the $t_{c1}$-$t_{c2}$ phase diagram
 obtained from the model of Fig.~\ref{et2}. 
 The ZGS state is realized in region I and 
 the insulating state is obtained in region IV, respectively. 
  The inset shows  the locations of contact points 
  in the $k_x$-$k_y$ plane, which are obtained from eq.~(\ref{k0xk0y}) 
  when  parameters $t_{c1}$-$t_{c2}$ are varied as ($C_t=-1,-2,-3$) 
\begin{align}
t_{c2}=t_{c1}+C_t \point 
\label{arrow2}
\end{align}
The corresponding parameters are 
 shown by  the arrow with  1, 2 and 3 in the  main frame. 
The boundary between  I and IV is given by 
\begin{align}
t_{c2}=-t_{c1}\pm 2 \virg 
\label{arrow2}
\end{align}
respectively. 
The properties of the contact point are as follows. 
 The contact point is located at X for $t_{c2}=-t_{c1}-2$, 
 and at $\Gamma$ for $t_{c2}=-t_{c1}+2$, 
   whereas the contact point is present 
    near the zone boundary of the $k_x$-$k_y$ plane
     for   $t_{c1} \simeq t_{c2}$.  
 When   $t_{c1}=t_{c2}$ holds, 
  two bands are degenerate on the zone boundary and  
    the contact point disappears.
These facts indicate that  
   the  contact  point  comes from  the symmetry breaking of the bond
      due to   $t_{c1} \neq t_{c2}$. 
      
From Fig.~\ref{c1c22site}, it is found that the ZGS state 
  is mainly obtained for  $t_{c1}  t_{c2} < 0$, 
 whereas the transfer energies  
 ($t_{p3}$, $t_{p4}$, $t_{c3}$, $t_{c4}$) are needed 
  for the metallic state.   
\section{Summary and Discussion}

In the present study, we examined the ZGS state by changing transfer energies 
 within the tight-binding model. 
For $\alpha$-(ET)$_2$I$_3$, the hole and electron bands are degenerate at two contact points 
$\mib{k}^0$ and $-\mib{k}^0$ in the first Brillouin zone. 
The dispersion of the electron and hole bands around the contact point 
resembles 
that of the massless Dirac Fermion, 
which is called the Dirac cone.
In a wide pressure range (also parameters of 
 transfer energy range), there exists the ZGS state, in which 
   the contact point coincides with the Fermi point. 
The location for the contact point moves in the $\mib{k}$ space continuously 
when transfer energies are varied. 
The stability of the contact point was examined using a simplified model. 
The contact point moves to 
 the symmetry axis  in the Brillouin zone
 when the parameters  are varied toward those with higher  symmetry.  
This was also verified using the reduced model.

Let us comment on the state under the $b$-axis pressure $P_b$, which is calculated from 
 eq.~(\ref{transfer_PB}). 
When  $P_{b}$ is increased, 
 the contact point still exists  and moves as shown  by 
  the arrow  $P_b$ $(<10)$  in Fig.~\ref{etfscp}.
 However, the ZGS state does not appear,   
 since the contact point is always located  below the Fermi energy.
 The absence of the ZGS state   comes from  the fact that 
  the  parameters $t_{c1}$ and $t_{c2}$  still 
  remain  in region II of Fig.~{\ref{c1c21}}.  
In other words, $t_{c1}$ for $P_b$ does not increase 
   much compared with that for  $P_a$.    

Let us examine if the ZGS state is present in other reduced models 
 for $\alpha$-(ET)$_2$I$_3$. 
Hotta \cite{hotta}
  examined a different model
  with two molecules in the unit cell 
  by taking  the dimer for the bond 
  $t_{p1}$   between  1 and 3 sites  
  (and also that for $t_{p2}$   between 2 and 4 sites) in 
   Fig.~\ref{unitcell},
  where  the five transfer energies are kept.  
 By substituting into  these energies, the experimental data at ambient pressure
 \cite{Kondo_2},  
  the contact point below the Fermi energy is obtained 
    at $\mib{k}^0 =(0.641\pi, -0.694\pi)$.  
   This  corresponds to the metallic state, 
     although the location of $\mib{k}^0$  differs from that   
  of  Fig.~\ref{band1}.
 For such a model, we also obtained the ZGS state under the pressure of $P_a$. 
 Comparing Hotta's model with the present one, 
  both models take into account the dimerization  for the bond with 
   the large transfer energy, 
  but there is a difference   
  in the choice of the bond as seen from Fig.~\ref{unitcell}. 
  These two kinds of pairings, which are introduced 
  for simplicity of the calculation,   seem to be almost degenerate 
   in their energy. 
 In the present model of  \S 5,   the  condition of  
  $t_{p3}=t_{p4}=t_{c3}=t_{c4}=0$   
   is further introduced   in order to  focus on the ZGS state. 
   There is also  another reduced model for $\alpha$-(ET)$_2$I$_3$, 
   \cite{Kino_JPSJ_ET2}
  in  which  $t_{c1} = t_{c2} =t_{c3} =t_{c4}$ is taken. 
   This model does not exhibit the ZGS state 
       within the numerical calculation. 
This result may be ascribed to the fact that 
   the ZGS state is expected mainly for $ t_{c1} t_{c2} <0$ in Fig.~6.  

Those contact points obtained in the present study are related to the 
"accidental degeneracy" of the bands, which was pointed out by Herring.\cite{Herring}
In the case of "accidental degeneracy", the degeneracy is not caused by the symmetry of 
the materials, but is caused by the intrinsic properties of the special band.
As for the general case of the three-dimensional system, there appear contact lines in the 
Brillouin zone, which lead to a metal due to the existence of the Fermi surface.
The ZGS state in $\alpha$-(ET)$_2$I$_3$ exhibits the contact point, that always coincides with the Fermi energy, 
due to the quarter-filling in the highly two-dimensional system.
The existence of such a degeneracy in $\alpha$-(ET)$_2$I$_3$ has been supported recently 
by first-principle calculations.\cite{Kino_Miyazaki,Ishibashi}

Anomalous properties associated with the Dirac cone have been investigated for bismuth\cite{Wolff,Kohno} 
and graphite\cite{Abrikosov,Ando_jpsj} using the effective Hamiltonian for the Dirac cone. 
In those cases the degeneracy occurs on the symmetry axis of its Brillouin zone.
The effect of Berry phase due to the Dirac cone has been observed 
using the phase analysis of quantum oscillations\cite{Lukyancuk}.
Thus, the ZGS state of $\alpha$-(ET)$_2$I$_3$ is also expected to exhibit many anomalous properties.
The new phenomenon known as anomalous transport
\cite{Kajita,Tajima_2000,Kajita_2003} may come from the Dirac cone and the unique properties of the ZGS state.

\section*{Acknowledgements}
The authors are thankful to  R. Kondo,  S. Kagoshima, 
 T. Takahashi and D. S. Hirashima  for useful comments. 
 The authors are 
   also grateful to S. Sugawara, N. Tajima, Y. Nishio and K. Kajita
  for useful  discussions.
      The present work has been financially supported 
      by a Grant-in-Aid for Scientific Research on Priority Areas of Molecular Conductors 
      (No. 15073103) from the Ministry of Education, Culture, Sports, Science and Technology, Japan.


\begin{thebibliography}{99}
\bibitem{Ishiguro_1998}
T. Ishiguro, K. Yamaji and G. Saito: {\it Organic Superconductors} 2nd ed. 
(Springer-Verlag, Berlin, 1998)
\bibitem{Hseo1}H. Seo, C. Hotta and H. Fukuyama: Chem. Rev. \textbf{104} (2004) 5005.
\bibitem{Bender} K. Bender, I. Hennig, D. Schweitzer, K. Dietz, H. Endres and H. J. Keller: 
Mol. Cryst. Liq. Cryst. \textbf{108} (1984) 359.
\bibitem{Rothaemel} B. Rothaemel, L. Forro, J. R. Cooper, J. S. Schilling, M. Weger, P. Bele, 
H. Brunner, D. Schweitzer and H. J. Keller\prb{34}{1986}{704}.
\bibitem{Kajita}
K. Kajita, T. Ojiro, H. Fujii, Y. Nishio, H. Kobayashi, A. Kobayashi and
R. Kato: J. Phys. Soc. Jpn. {\bf 61} (1992) 23. 
\bibitem{Tajima_2000}N. Tajima, M. Tamura, Y. Nishio, K. Kajita and 
Y. Iye\jpsj{69}{2000}{543}.
\bibitem{Takano}
Y. Takano, K. Hiraki, H. M. Yamamoto, T. Nakamura and T. Takahashi: J. Phys. Chem. Solids \textbf{62} (2001) 393.
\bibitem{Wojciechowskii}
R. Wojciechowskii, K. Yamamoto, K. Yakushi, M. Inokuchi and A. Kawamoto\prb{67}{2003}{224105}.
\bibitem{Takahashi_SM_2003}
T. Takahashi:
Synth. Met. {\bf 133-134} (2003) 261.
\bibitem{Maesato}
 M. Maesato, Y. Kaga, R. Kondo and S. Kagoshima:
  Rev. Sci. Instrum. {\bf 71}  (2000) 176. 
\bibitem{Tajima_2002}N. Tajima, A. Ebina-Tajima, M. Tamura, Y. Nishio 
and K. Kajita\jpsj{71}{2002}{1832}.
\bibitem{Kino_JPSJ_ET}
H. Kino and H. Fukuyama:
J. Phys. Soc. Jpn. {\bf 64} (1995) 4523.
\bibitem{Kino_JPSJ_ET2}
H. Kino and H. Fukuyama:
J. Phys. Soc. Jpn. {\bf 65} (1996) 2158.
\bibitem{Seo_JPSJ}
H. Seo\jpsj{69}{2000}{805}.
\bibitem{hotta} 
C. Hotta: J. Phys. Soc. Jpn. {\bf 72} (2003) 840. 
\bibitem{Kobayashi_JPSJ}
A. Kobayashi, S. Katayama, K. Noguchi and Y. Suzumura: 
 J. Phys. Soc. Jpn. {\bf 73} (2004) 3135. 
\bibitem{Kobayashi_JPSJ2}
A. Kobayashi, S. Katayama  and Y. Suzumura: 
J. Phys. Soc. Jpn. {\bf 74} (2005) 2897. 
\bibitem{Mori}
T. Mori, H. Mori and S. Tanaka: 
Bull. Chem. Soc. Jpn. {\bf 72} (1999) 179.
\bibitem{Kondo}
   R. Kondo and S. Kagoshima: 
    J. Phys. IV France {\bf 114} (2004) 523. 
\bibitem{Kondo_2}
   R. Kondo, S. Kagoshima and J. Harada: 
    Rev. Sci. Instrum. {\bf 76} (2005) 093902.
\bibitem{Herring}
C. Herring: Phys. Rev. {\bf 52} (1937) 365.
\bibitem{Kohno}
H. Kohno, H. Yoshioka and H. Fukuyama: J. Phys. Soc. Jpn. {\bf 61} (1992) 3462. 
\bibitem{Wolff}
P. A. Wolff: J. Phys. Chem. Solids {\bf 25} (1964) 1057.
\bibitem{Abrikosov}
A. A. Abrikosov: Phys. Rev. B {\bf 60} (1999) 4231.
\bibitem{Lukyancuk}
I. A. Luk'yanchuk and Y. Kopelevich: Phys. Rev. Lett. {\bf 93} (2004) 166402.
\bibitem{Sugawara}
Another  model for the ZGS state has also been discussed. 
(S. Sugawara: Master Thesis, Toho University 2005);  
S. Sugawara, N. Tajima, Y. Nishio and K. Kajita: private communication.  
\bibitem{Kino_Miyazaki} 
H. Kino and T. Miyazaki: J. Phys. Soc. Jpn. {\bf 75} (2006) 034704. 
\bibitem{Ishibashi}
S. Ishibashi, T. Tamura, M. Kohyama and K. Terakura: 
J. Phys. Soc. Jpn. {\bf 75} (2006) 015005. 
\bibitem{Ando_jpsj}
T. Ando: J. Phys. Soc. Jpn. {\bf 74} (2005) 777. 
\bibitem{Kajita_2003}
K. Kajita, N. Tajima, A. Ebina-Tajima and Y. Nishio: Synth. Met. {\bf 133-134} (2003) 95.
\end{thebibliography}
\end{document}